\documentclass[
prd,aps,
showpacs,tightenlines,nofootinbib,preprintnumbers, superscriptaddress
]{revtex4}
\usepackage{graphicx}
\usepackage{amssymb,amsmath,latexsym}
\usepackage{amsfonts}
\usepackage{bm}
\usepackage{color}

\newcommand{\beqa}{\begin{eqnarray}}
\newcommand{\eeqa}{\end{eqnarray}}

\newcommand{\siml}{\lesssim}

\newcommand{\beq}{\begin{equation}}  \newcommand{\eeq}{\end{equation}}
\newcommand{\bef}{\begin{figure}}  \newcommand{\eef}{\end{figure}}
\newcommand{\bec}{\begin{center}}  \newcommand{\eec}{\end{center}}

\newcommand {\ga} {\ {\raise-.5ex\hbox{$\buildrel>\over\sim$}}\ }
\newcommand {\la} {\ {\raise-.5ex\hbox{$\buildrel<\over\sim$}}\ }

\begin{document}

\title{Shadows of Multi-Black Holes: Analytic Exploration}

\author{Akifumi Yumoto}
\affiliation{Department of Physics, 
College of Humanities and Sciences, 
Nihon University, 
Tokyo 156-8550, Japan}
\author{Daisuke Nitta}
\affiliation{Astronomical Institute, Tohoku University, Aoba Aramaki Aoba, Sendai 980-8578, Japan}
\author{Takeshi Chiba}
\affiliation{Department of Physics, 
College of Humanities and Sciences, 
Nihon University, 
Tokyo 156-8550, Japan}
\author{Naoshi Sugiyama} 
\affiliation{Department of Physics, Nagoya
  University, Chikusa, Nagoya 464-8602, Japan} \affiliation{Kavli Institute
  for Physics and Mathematics of the Universe, University of Tokyo,
  Chiba 277-8582, Japan}

\date{\today}

\pacs{97.60.Lf  ; 04.70.-s }

\begin{abstract}
Shadows of multi-black holes have structures distinct from the mere superposition of the shadow of a single black hole:  the eyebrow-like structures outside the main shadows and the deformation of the shadows. We present analytic estimates of these structures using the static multi-black hole solution (Majumdar-Papapetrou solution). 
We show that the width of the eyebrow is related with the distance between 
the black holes and that the shadows are deformed into ellipses due to the presence  of the second black holes. These results are helpful to understand 
qualitatively the features  
of the shadows of colliding black holes. We also present the shadows of colliding/coalescing black holes in the Kastor-Traschen solution. 
\end{abstract}

\maketitle

\section{Introduction}

One of the hottest topics in galaxy formation is the coevolution of 
super massive black holes (SMBHs) with spheroid components (bulges) of galaxies.   
It becomes more and more clear that most of galaxies and AGNs have at least one SMBH and 
there is a strong correlation between the SMBH mass 
and the bulge mass of host galaxies~\cite{kormendy,magorrian,merritt}.   
Although a detailed mechanism of the coevolution is not yet understood,  
it is almost certain galaxy mergers play an essential role, since it is known that 
bulges or spheroid components are formed by merger of galaxies 
in the hierarchical clustering scenario of structure formation.   
Hence it may be also natural to consider formation of SMBHs 
is due to mergers of smaller black holes.  

An observation clue of existence of binary black holes is recently 
obtained from detailed study of Kepler motion of a radio emission component 
in the radio galaxy 3C 66B by using a
technique of phase-referencing very-long-baseline interferometry (VLBI)~\cite{sudou}. 
In particular,  a newly found periodic flux variation suggests that this binary system will coalesce in 
500 years~\cite{iguchi}.  
Perhaps, we may conclude coalescence of binary black holes often takes place 
in the Universe.   However a direct evidence of black hole merger is still missing.  
One of the possibilities to {\it see} the merger process is to observe shadows of black holes 
shone by the radiation from the accretion disc or star lights behind the black holes~\cite{falcke}. 
Since two event horizons merge into one event horizon, we expect that the shadows must show 
very peculiar time evolution.   Observing these shadows, therefore,  should be compelling evidence of 
a coalescing black holes as well as provides an intriguing probe of general relativity with very strong 
gravity field.  

As a first step toward the study of a realistic black hole binary, we have 
recently calculated  the shadows of the Kastor-Traschen (KT) \cite{kt} cosmological 
multi-black hole solutions \cite{nitta}. 
We have found that the shadows are deformed in the direction of the collision 
and that in addition to the shadow of each black hole, eyebrow-like structures appear as the black holes come close to each other. 
In this paper, we attempt to understand these structures analytically using 
the Majumdar-Papapetrou (MP) solution \cite{mp}, the  static multi-black hole solution with the maximal charge to which the KT 
solution is reduced when the cosmological constant is zero.

The paper is organized as follows. 
In Sec. II, we present several analytic calculations of null geodesics in 
the extreme Reissner-Nordstr\"om and MP solutions in order 
to provide  analytic estimate of the eyebrow-like 
structures of the shadows as well as the deformation of the shadows  
in the MP solution. 
In Sec. III, extending our previous results \cite{nitta}, we present 
numerical results of the shadows of 
the colliding/coalescing  black hole binary in the KT solution by changing the viewing angle of the distant observer and {compare them with the shadows 
of the MP solution.} 
Sec. IV summarizes the results.

\section{Shadows of Static Two Black Holes: Majumdar-Papapetrou Solution}

\subsection{Geodesics in Extreme Reissner-Nordstr\"om solution}

The Majumdar-Papapetrou (MP) solution \cite{mp} is given by
\beqa
&&ds^2=-\Omega^{-2}dt^2+\Omega^2(dx^2+dy^2+dz^2),\label{eq:mp}\\
&&\Omega=1+\sum_{i}\frac{m_i}{r_i},\quad r_i\equiv \sqrt{(x-x_i)^2+(y-y_i)^2+(z-z_i)^2},\nonumber
\eeqa
where $m_i$ is the mass of $i$-th black hole located at $r=r_i$ (we use the geometrical units, $G=c=1$). 
The MP solution in the case of a single black hole is reduced to the extremal 
Reissner-Nordstr\"om (RN) solution. 

First of all, we derive the relation between the deflection angle and the impact parameter close to the unstable circular orbit for the extreme RN solution 
following Luminet \cite{luminet,chandra}, in order to understand the shape of black hole shadows in MP spacetime. 

\begin{figure}
  \begin{center}
   \includegraphics[width=11cm]{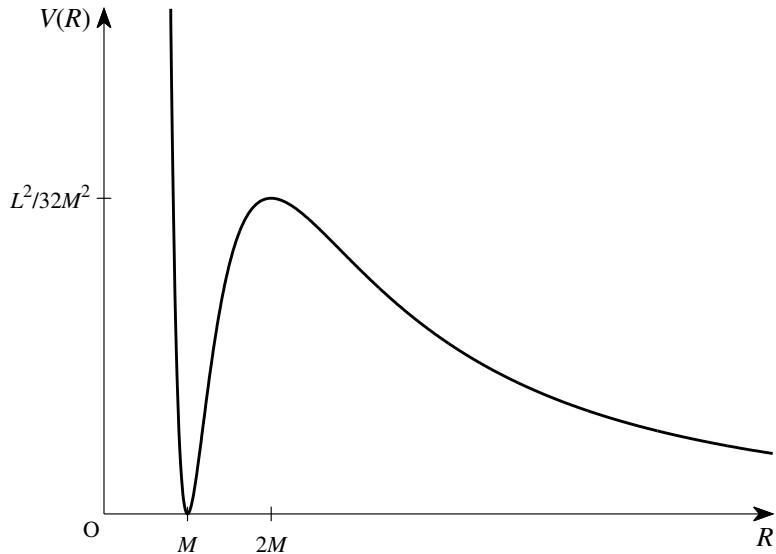}
  \end{center}
  \caption{The effective potential $V(R)$ for the extreme RN black hole. 
  \label{fig:RN-Q=M_V-r.pdf}}
\end{figure}

The metric of the extreme RN solution is given by
\beqa
ds^2 = -\left( 1 - \frac{M}{R} \right)^2 dt^2
+ \left( 1 - \frac{M}{R} \right)^{-2} dR^2
+ R^2 \left( d\theta^2 + \sin^2 \theta d\phi^2 \right),
\eeqa
and the horizon is at $R=M$. The radial coordinate $R$ is related with the isotropic coordinate $r$ in Eq. (\ref{eq:mp}) as $R=r+M$. 
{}From the spherical symmetry, we may restrict ourselves to the equatorial plane without loss of generality. 
In terms of two conserved quantities, i.e. 
the energy $E = \left( 1 - \dfrac{M}{R} \right)^{2} \dot{t}$ and the angular momentum $L = R^2 \dot{\phi}$, 
the null geodesics satisfy the "energy equation"
\beqa
&&\frac12{\dot{R}^2} + V(R)
= \frac12{E^2},\label{eq:RN-Q=M_energy}\\ 
&&V(R)=\frac{L^2}{2R^2} \left( 1 - \frac{M}{R} \right)^{2},
\nonumber
\eeqa 
where $\dot t=dt/d\lambda$ with $\lambda$ being the affine parameter. 
$V(R)$ is shown in Fig. \ref{fig:RN-Q=M_V-r.pdf}.  
The null geodesics with $E^2/2$ being larger than 
$L^2/32M^2$, which is  
the local maximum of $V(R)$ at $R=2M$,  fall into the black hole. 
In terms of the impact parameter 
$b=L/E$, the light rays with $b<4M$ fall into the black hole. 
Thus, the shadow of the extreme RN black hole is the disk of radius $4M$. 

Let us investigate  the behaviors of the null geodesics approaching the unstable circular orbit at $R=2M$. 
In terms of $u \equiv 1/R$, Eq. (\ref{eq:RN-Q=M_energy}) is written as
\beqa
\left( \frac{du}{d\phi} \right)^2
= -u^2 \left( 1 - Mu \right)^2 + \frac{1}{b^2}\equiv f(u).
\label{eq:dudphi}
\eeqa
For $b > 4M$, the quartic equation $f(u)=0$ has four distinct roots, 
three of which are positive.  
Arranging the ordering of the roots as $u_1 < 0< u_2 < u_3 < u_4$,  
$1/u_2$ corresponds to the periastron distance. 
Then the solution of Eq. (\ref{eq:dudphi}) can be written as
\beqa
\phi_{\infty} &=& \int_0^{u_2} \frac{du}
{\sqrt{\left\{ \left( \dfrac{1}{4M} + \dfrac{1}{b} \right) - M \left( \dfrac{1}{2M} - u \right)^2 \right\}
\left\{ M \left( \dfrac{1}{2M} - u \right)^2 - \left( \dfrac{1}{4M} - \dfrac{1}{b} \right) \right\}}} \nonumber\\
&=&
2\sqrt{\frac{b}{b+4M}} \left[ K(k) 
- F \left( \frac{\pi}{4}, k \right) \right],
\eeqa
where the origin of $\phi$ is chosen at the periastron passage when $u=u_2$ and 
$k$ is given by 
\beqa
k^2=\frac{8M}{b+4M}. 
\eeqa
$F(\varphi, k)$ is  the elliptic integral of the first kind 
defined by
\beqa
F(\varphi, k) = \int_0^{\varphi} \dfrac{d\theta}{\sqrt{1 - k^2 \sin^2 \theta}},
\eeqa
and $K(k)=F(\pi/2,k)$ is the complete elliptic integral of the first kind.  

{}The solution can be used to obtain the asymptotic behavior of $\phi_{\infty}$ for 
$1/u_2 \rightarrow 2M$. If we write  $1/u_2=2M(1 + \delta)$ with $\delta\ll 1$, 
then from $f(u_2)=0$ in Eq. (\ref{eq:dudphi}),  $b=4M(1+\delta^2)$ and hence $k^2=1-\frac{\delta^2}{2}$. Using the asymptotic relation $K(k)\rightarrow \frac12\ln \left({16}/(1-k^2)\right)$ 
for $k\rightarrow 1$, we obtain 
\beqa
\phi_{\infty} = \frac{1}{\sqrt{2}}\ln \frac{2^5}{(\sqrt{2}+1)^2\delta^2}.
\eeqa
Therefore, the light rays with the impact parameter $b=4M(1+\delta^2)$ are deflected by the black hole by the angle $\Theta=2\phi_{\infty} -\pi $  
with the relation  
\beqa
b=4M + \frac{2^7 M}{\left( \sqrt{2} + 1 \right)^2} e^{-\frac{\pi}{\sqrt{2}}} e^{-\frac{1}{\sqrt{2}} \Theta}.
\label{eq:b-Theta}
\eeqa
For those geodesics that go round the black hole $n$-times,  
the deflection angle can be written as $\Theta +2n\pi$.  
Accordingly, we have 
\beqa
b=4M + \frac{2^7 M}{\left( \sqrt{2} + 1 \right)^2} e^{-\frac{\pi}{\sqrt{2}}} e^{-\frac{1}{\sqrt{2}} (\Theta + 2n\pi)}.
\eeqa

\subsection{Shadows of MP solution}

\begin{figure*}[htbp]
\includegraphics[width=13cm]{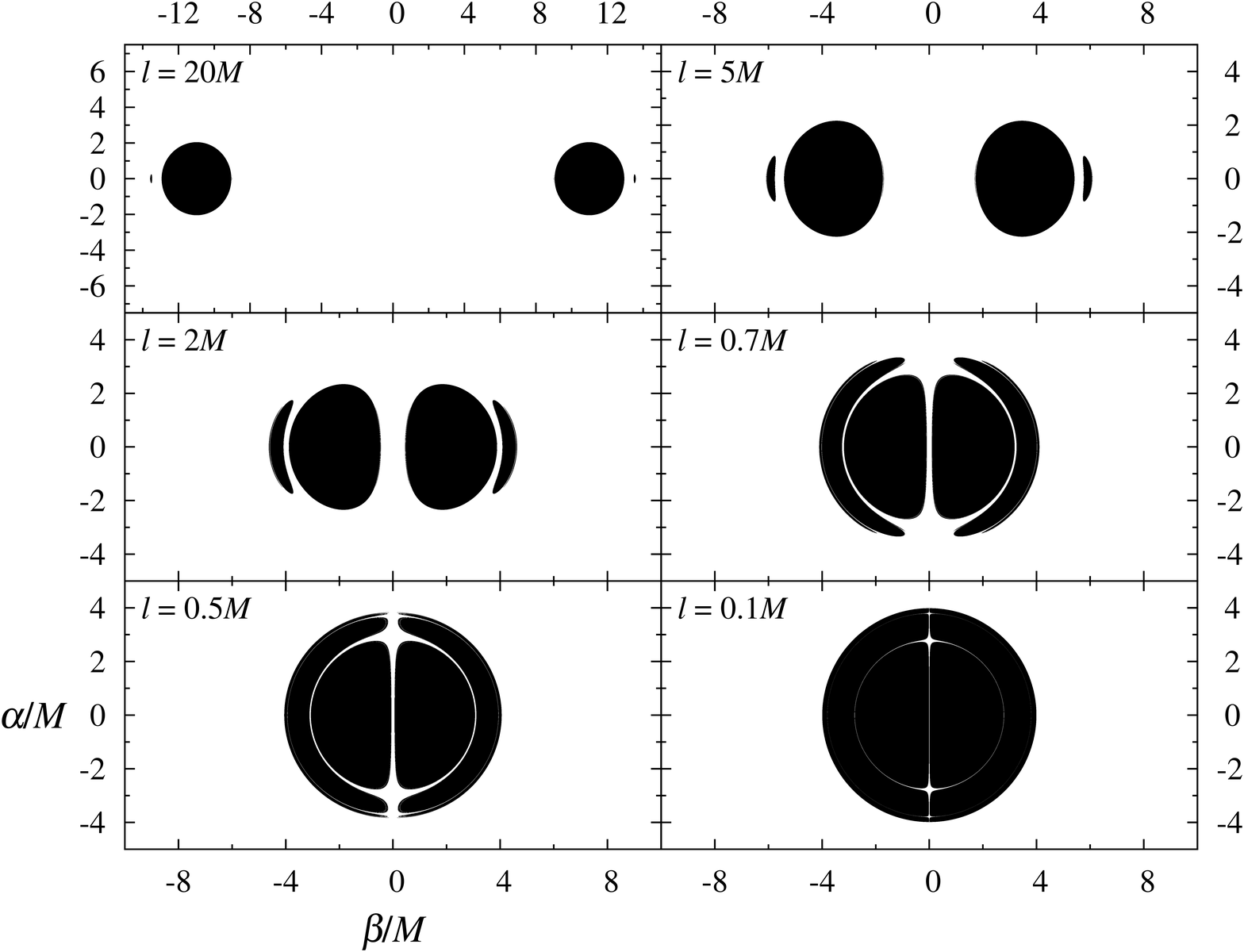}
\caption{Shadows of two black holes in the MP spacetime for several distances 
$\ell$ between the black holes. 
\label{fig:MP-shd.pdf}}
\end{figure*}

Next, we study the shadows of two black holes in the MP spacetime. 
We consider a two equal mass black hole system in the MP spacetime 
Eq. (\ref{eq:mp}). 
The black holes are located at $z=\pm \ell/2$ with mass $m=M/2$. Because of 
the axial symmetry, there exists a conserved quantity $L_z$ 
(corresponding to the angular momentum integral) for the geodesics 
as well as the energy integral which are defined in the spherical coordinate by
\beqa
E=\Omega^{-2}\dot t, \quad  L_z=\Omega^2r^2\sin^2\theta \dot\phi.
\label{integral}
\eeqa

We calculate the null geodesics from the observer located far from 
the black holes. The geodesics that fall into the black hole form the shadow 
of the black holes. 
In Fig. \ref{fig:MP-shd.pdf}, the shadows of two black holes in the MP spacetime are shown. The observer is located at the equatorial plane ($\theta=\pi/2$). 
Here we have defined the celestial coordinate of an observer, $(\alpha,\beta)$, as 
\beqa
\alpha=\lim_{r\rightarrow \infty} -\frac{rP^{(\phi)}}{P^{(t)}},  \quad 
\beta=\lim_{r\rightarrow \infty}\frac{rP^{(\theta)}}{P^{(t)}},
\label{celestial}
\eeqa
where $P^{(\mu)}$ are the momenta in the local inertial frame. 

We find that there appears additional eyebrow-like structure in the outer 
region of the main shadow. The eyebrow grows as the distance 
between the black holes is closer. Although not discernible in the figure,  
 in fact there appears "the fine structure" of the eyebrows:  
infinitely many thinner eyebrows 
at the outer region of these eyebrows as well as at the inner region of the main shadow. 

Moreover, the distance between the main shadows is slightly larger than 
the real distance between the black holes, and  
the shapes of the shadows are suppressed in the $\beta$ direction and are slightly elongated in the $\alpha$ direction. 
Note that the radius of the shadow of a single black hole is $4m_i$.  
Take $\ell=2M$ case as an example. If the other black hole is absent, the shadow should be located at $\beta/M=\pm 1$ with the radius $4m=2M$. However, because of the presence of the second black hole, 
the inner edge of the shadow is shaved and the outer edge is stretched toward larger $\beta$ instead. 
In the following we study how these structures, different from the superposition 
of the shadow of a single black hole, appears.

\subsection{Eyebrow in Black Hole Shadow}

\begin{figure}[htbp]
\includegraphics[width=9.5cm]{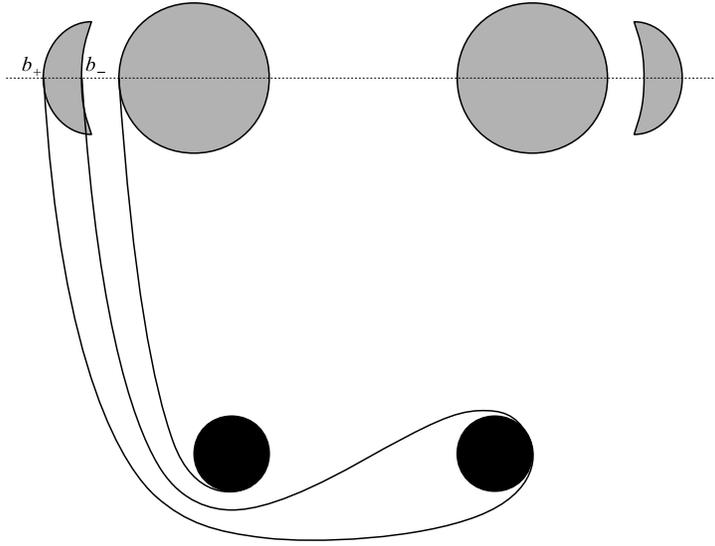}
\caption{Null geodesics for the "eyebrow" of black hole shadows. 
The geodesics coming in with the impact parameter $b_+$ and $b_- (<b_+)$ are deflected 
by the left black hole.  Then they go into the unstable circular orbit of 
the right black hole: the one with $b_+$ from the outside, while inside with $b_-$. 
The geodesics with the impact parameters between $b_-$ and $b_+$ form the eyebrow of the shadow. 
\label{fig:MP_db-l_fig-schm.pdf}}
\end{figure}

First, we try to explain how the "eyebrows" are formed. 
We restrict ourselves to the $\alpha=0$ plane. The schematic picture of null 
geodesics for the silhouette of the eyebrows is shown in Fig. \ref{fig:MP_db-l_fig-schm.pdf}. 
The geodesics coming in with impact parameter $b_+$ and $b_- (<b_+)$ are deflected 
by the left black hole.  The geodesics with $b_+$ go into the unstable circular 
orbit of the right black hole from the outside, while inside for those with $b_-$. 
The geodesics with the impact parameters between $b_-$ and $b_+$ form the eyebrow of the shadow. Hence, the width of the eyebrow $\Delta b = b_+ - b_-$ is directly related with the distance between the hole $\ell$.\footnote{Likewise, the fine structure of the eyebrows may be understood as those null geodesics that are 
captured by the one hole after going around the other hole several times.}

\begin{figure}[htbp]
 \begin{minipage}{0.52\hsize}
  \begin{center}
   \includegraphics[width=0.9\hsize]{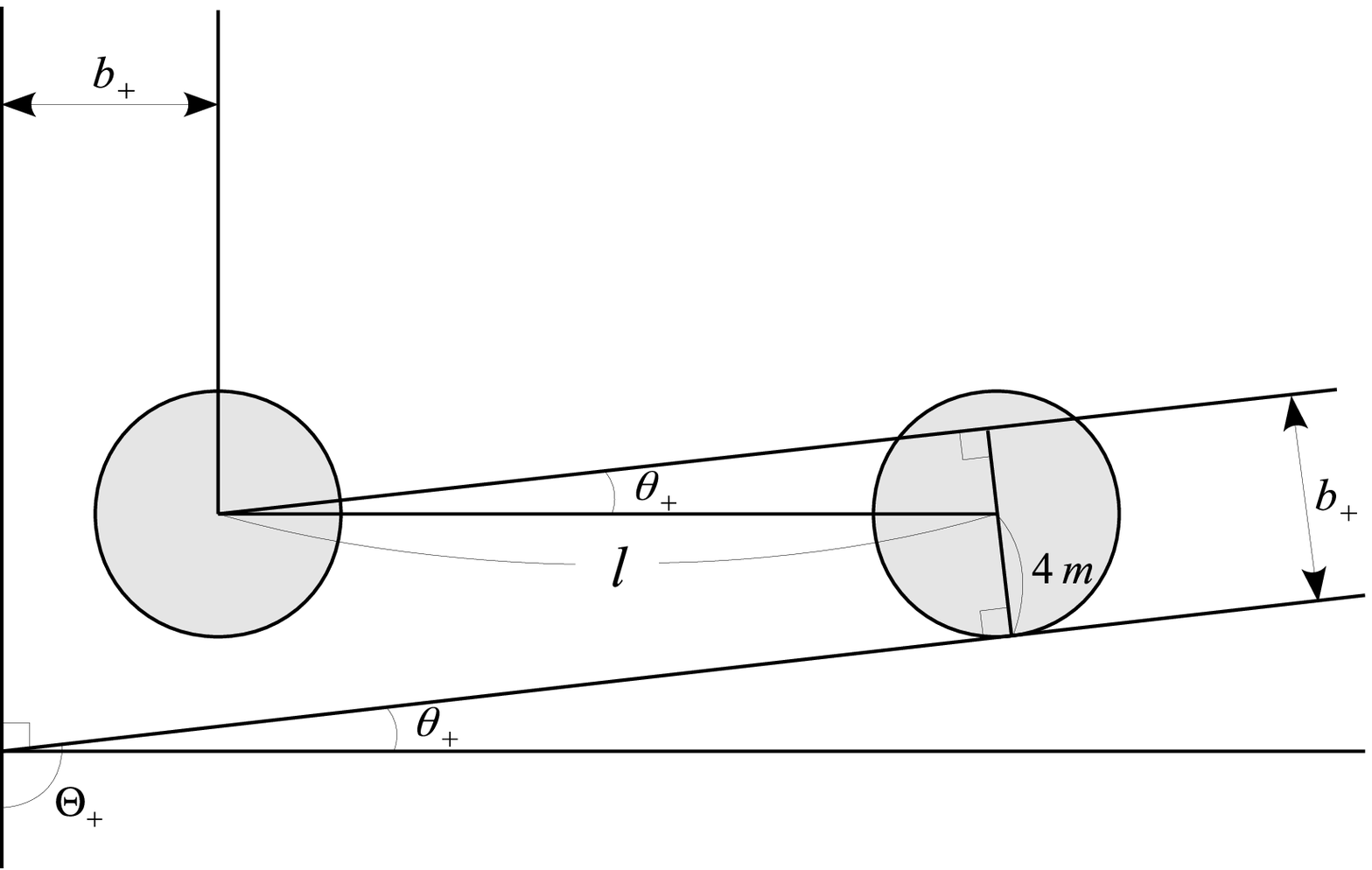}
  \end{center}
  \caption{The relation between $b_+$, $\Theta_+$ and $\ell$.
  \label{fig:MP_db-l_fig-bpls.pdf}}
 \end{minipage}%
 \begin{minipage}{0.48\hsize}
 \begin{center}
  \includegraphics[width=0.9\hsize]{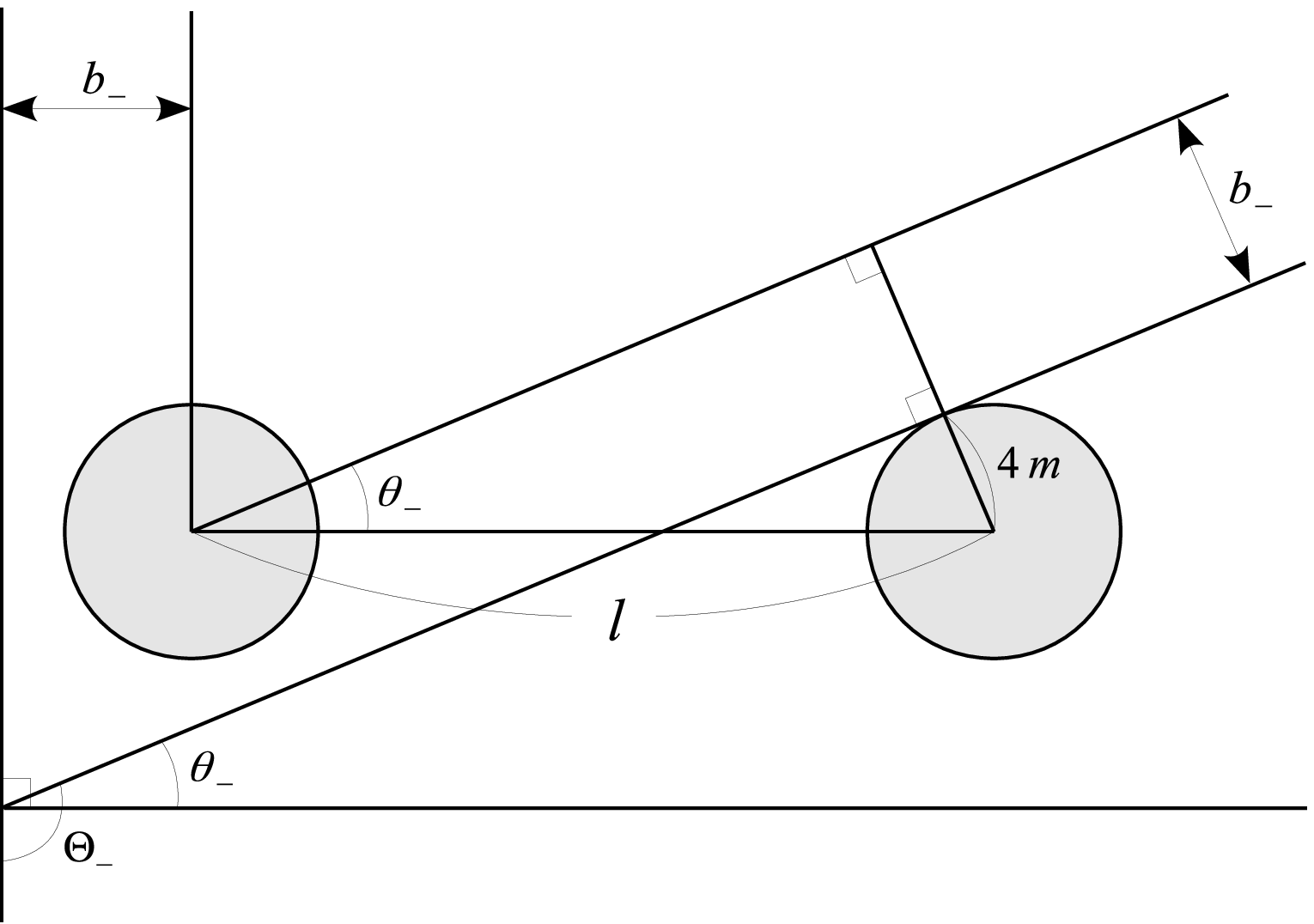}
 \end{center}
  \caption{The relation between $b_-$, $\Theta_-$ and $\ell$.
  \label{fig:MP_db-l_fig-bmns.pdf}}
 \end{minipage}
\end{figure}

Then we calculate the relation between the deflection angle and the impact parameter and the distance.  We assume the distance is so large $\ell \gg m$ that the effect of the second black hole on the spacetime may be treated 
as a perturbation to a single black hole. 
The mass of the black holes is assumed to be equal with $m=M/2$. 
{}From Fig. \ref{fig:MP_db-l_fig-bpls.pdf} 
and Fig. \ref{fig:MP_db-l_fig-bmns.pdf}, we find that 
the deflection angle $\Theta_{\pm}$  of the geodesics with the impact parameter $b_{\pm}$ is given by
\beqa
\Theta_+ = \frac{\pi}{2} + \frac{b_+ - 4m}{\ell},  \quad  
\Theta_- = \frac{\pi}{2} + \frac{b_- + 4m}{\ell}.
\eeqa
{}Then from Eq. (\ref{eq:b-Theta})
\beqa
\frac{\Delta b}{m} = \frac{b_+ - b_-}{m}
= \frac{2^7}{\left( \sqrt{2} + 1 \right)^2} e^{-\frac{\pi}{\sqrt{2}}}
\left( e^{-\frac{\Theta_+}{\sqrt{2}}} - e^{-\frac{\Theta_-}{\sqrt{2}}} \right) 
=\frac{2^7 m}{\left( \sqrt{2} + 1 \right)^2 \sqrt{2} \,\ell} e^{-\frac{3\pi}{2\sqrt{2}}}
\left( -\frac{\Delta b}{m} + 8 \right).
\eeqa
Hence we obtain to the first order in $m/\ell$
\beqa
\Delta b = \dfrac{2^{10} m^2}{\left( \sqrt{2} + 1 \right)^2 \sqrt{2} \,\ell} e^{-\frac{3\pi}{2\sqrt{2}}}\simeq 1.1\frac{M^2}{\ell}.
\label{eq:db(l)}
\eeqa

\begin{figure}[htbp]
\includegraphics[width=11cm]{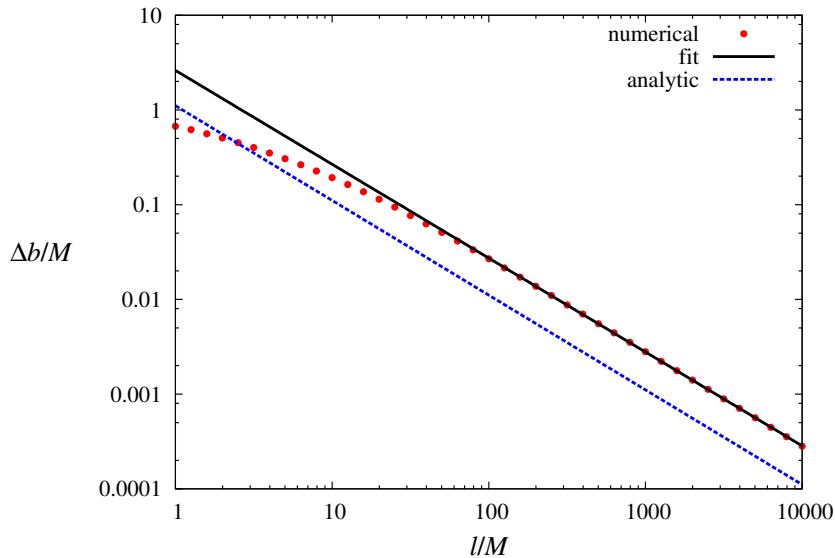}
\caption{The width of the eyebrow $\Delta b$ as a function of the distance between the holes. The numerical results (dot: red) and its linear fit (solid: black).   
The dashed (blue) line is the analytic relation Eq. (\ref{eq:db(l)}). 
\label{fig:MP_db-l_log,fit(lin)10-1000.pdf}}
\end{figure}

We numerically compute the width of the eyebrow $\Delta b$ for several $\ell$. 
The results are shown in Fig. \ref{fig:MP_db-l_log,fit(lin)10-1000.pdf}. 
The least square fit of the numerical data for $10^2 \leq \ell/M \leq 10^4$ gives 
\beqa
\ln \frac{\Delta b}{M} = 0.96 -0.99 \ln \left( \frac{\ell}{M} \right),
\eeqa
or $\Delta b/M=2.6(\ell/M)^{-0.99}$. Apart from the slight offset, the power 
index  is in good agreement with the analytic estimate Eq. (\ref{eq:db(l)}).

\subsection{Deformation of Black Hole Shadow}

Next, we study how the shadows are deformed and are shifted 
toward a larger separation. 

We assume that two black holes are located at $z=0$ and $z=\ell$.  
In the spherical coordinate,  $\Omega$ is given by
\beqa
\Omega=1+\frac{m}{r}+\frac{m'}{r'}\equiv 1+\frac{m}{r}+\psi(r,\theta),\quad 
r'=\sqrt{r^2+\ell^2-2\ell r\cos\theta}.
\label{omega}
\eeqa
We assume $\ell \gg m,m'$, in order to treat the effect of the black hole $m'$ 
at $z=\ell$ as a perturbation to the black hole $m$ at $z=0$.  


In addition to the conserved quantities $E$ and $L_z$ (Eq. (\ref{integral})), 
we also define the quantity $Q$ which corresponds to the Carter constant for 
a single black hole
\beqa
Q\equiv P_{\theta}^2+\cot^2\theta L_z^2=L_x^2+L_y^2,
\eeqa
where $P_{\theta}=\Omega^2r^2\dot\theta$. Introducing 
$\zeta^2\equiv Q/E^2$, $\xi\equiv L_z/E$, we have for the null geodesics
\beqa
\frac{1}{E^2}\dot r^2=1-\frac{\zeta^2+\xi^2}{\Omega^4r^2}\equiv {\cal R}(r).
\label{eq:dr}
\eeqa

Note that $\zeta$ (or $Q$) is no longer conserved due to the presence 
of the black hole $m'$. In fact, from the null geodesic equation for $\theta$, we have
\beqa
\frac{d\zeta^2}{d\lambda}=4\Omega\Omega_{,\theta}P_{\theta}.
\label{eq:zeta}
\eeqa
Introducing $\mu\equiv \cos\theta$, then 
in terms of the derivative with respect to $r$, Eq. (\ref{eq:dr}) and Eq. (\ref{eq:zeta}) can be written as
\beqa
&&\left(\frac{d\mu}{dr}\right)^2=\frac{\zeta^2-(\zeta^2+\xi^2)\mu^2}
{r^4\Omega^4-(\zeta^2+\xi^2)r^2}
,\label{eq:dmu}\\
&&\frac{d\zeta^2}{dr}=4r^2\Omega^3\Omega_{,\mu}\frac{d\mu}{dr}
\label{eq:dzeta}.
\eeqa

The equations determining the unstable orbits are 
\beqa
&&{\cal R}=1-\frac{\zeta^2+\xi^2}{\Omega^4r^2}=0\label{r=0}\\
&&\frac{d{\cal R}}{dr}=
-\frac{1}{r^2\Omega^4}\frac{d\zeta^2}{dr}+\frac{\zeta^2+\xi^2}{r^4\Omega^8}\frac{d}{dr}(r^2\Omega^4)=0.
\label{dr=0}
\eeqa
Eq. (\ref{r=0}), Eq. (\ref{dr=0}) and Eq. (\ref{eq:dzeta}) are combined to give
\beqa
&&\zeta^2+\xi^2=r^2\Omega^4,\label{eq:V}\\
&&\Omega+2r\Omega_{,r}=0.\label{eq:dV}
\eeqa
Denoting the solution to the above equations as $r_c$, $r_c$ describes  
the shape of the photon sphere deformed by the second black hole. 

We set $m=1$ and $m'=\kappa$ and measure the length in units of $m$ for 
notational simplicity. 
Then for a single black hole, the radius of the photon sphere $r_c$ is $r_c=1$ and 
the radius of the black hole shadow is $4$. 

We expand $\zeta$ according to the order of $1/\ell$ for fixed $r$,
\beqa
\zeta=\zeta_0+\zeta_1+\zeta_2+\ldots .
\eeqa
The radius of the photon sphere $r_c$ is also expanded
\beqa
r_c=1+\epsilon_1+\epsilon_2+\ldots .
\eeqa
We consider the observer located at $\theta \sim \pi/2$. Then $r\mu/\ell\ll 1$ holds everywhere on the unstable orbit. So
$\psi$ in Eq. (\ref{omega}) can be expanded in terms of $\mu$
\beqa
\psi(r,\mu)=\psi_0(r)+\psi_1(r)\mu+\frac{1}{2}\psi_2(r)\mu^2+
\frac{1}{3!}\psi_3(r)\mu^3+
\ldots,
\eeqa
where
\beqa
\psi_0=\frac{\kappa}{\sqrt{r^2+\ell^2}},\quad
\psi_1=\psi_0\frac{\ell r}{r^2+\ell^2},\quad
\psi_n=\psi_0(2n-1)!!\left(\frac{\ell r}{r^2+\ell^2}\right)^n.
\eeqa
For the observer located at $(r,\theta,\phi)=\lim_{r\to \infty}(r,\pi/2,0)$, 
$\xi$ and $\zeta$ are related to the celestial coordinate $\alpha$ and $\beta$ 
(Eq. (\ref{celestial})) as 
$\xi=\alpha$, $\zeta_0(r\to \infty)=\beta$ \cite{chandra}. 

As explained in Appendix \ref{app1}, it is shown that up to ${\cal O}(1/\ell)$, $\zeta$ satisfies the following equation
\beqa
\frac{d\zeta}{dr}=-2\Omega\psi_{,\mu}. 
\label{eq:dzeta01}
\eeqa

\subsubsection{0th order}

{} From Eq. (\ref{eq:dzeta01}), $\zeta_0$ satisfies 
\beqa
\frac{d\zeta_0}{dr}=- 2\psi_1,
\label{eq:dzeta0}
\eeqa
where we have neglected $1/r$ as a small perturbation. 
Since $\zeta_0\to \beta$ as $r\to \infty$, the solution is
\beqa
\zeta_0=\beta+\frac{2\kappa \ell}{\sqrt{r^2+\ell^2}}.
\eeqa
We need to determine $r_c$ to calculate $\zeta_0$. From Eq.(\ref{eq:dV}), 
 $r_c=1$ to the zero-th order and then $\zeta_0=\beta+2\kappa$. 
Hence using Eq.(\ref{eq:V}), we find 
\beqa
\alpha^2+(\beta+2\kappa)^2=16.
\eeqa
In the zero-th order, the shape of the black hole shadow is a circle
 with the radius of $4$ which is the same as a single black hole. 
However the center is shifted toward $\beta<0$ direction by $-2\kappa$. 
This explain why the distance between the main shadow is slightly larger than 
the real distance between the black holes. 
The reason is simple: the null geodesics with $\beta>0$ are attracted not only 
by the black hole at $z=0$ but also by the black hole at  $z=\ell$ 
in the opposite direction, so they are deflected less, while for $\beta<0$ the null 
geodesics are attracted by both black holes in the same direction, so 
they are deflected more. Hence, this feature should not be limited to the 
MP spacetime, but should be present for uncharged black holes.

\subsubsection{1st order}

{}From Eq. (\ref{eq:dzeta01}), $\zeta_1$ satisfies 
\beqa
\frac{d\zeta_1}{dr}=
\frac{d\zeta_0}{dr}\left(\frac{1}{r}+\psi_0+\frac{\psi_2}{\psi_1}\mu\right). 
\label{eq:dzeta1}
\eeqa
In terms of $\chi$ defined by $r=\ell\tan\chi$, it is rewritten as 
\beqa
\frac{d\zeta_1}{d\chi}
=-\frac{2\kappa}{\ell}
(\cos\chi+\kappa\sin\chi\cos\chi
+3\ell\mu\sin^2\chi\cos\chi).
\eeqa
The integration of the third term is
\beqa
3\ell\int_{\pi/2}^{\chi} \mu\sin^2\chi\cos\chi d\chi
&=&\ell\mu\sin^3\chi-\ell\int_{\pi/2}^{\chi}\sin^3\chi\frac{d\mu}{d\chi}d\chi,
\nonumber\\
&=&\ell\mu\sin^3\chi+
\int_{\pi/2}^{\chi}\zeta_0 h\sin\chi d\chi\nonumber\\
&\simeq & \ell\mu\sin^3\chi
-\zeta_0\cos\chi
\eeqa
where we have set $h=1$ as explained in Appendix \ref{app1}. Hence the solution is
\beqa
\zeta_1=\frac{2\kappa}{\ell}(1-\sin\chi)-\frac{\kappa^2}{\ell}\cos^2\chi
+\frac{2\kappa}{\ell}\zeta_0\cos\chi-2\kappa\mu\sin^3\chi. 
\eeqa
Since the radius of the photon sphere is $r_c\simeq 1$, $\chi\simeq r/\ell$ and 
the last term can be neglected. Hence we have
\beqa
\zeta_1=(2\beta+3\kappa+2)\delta,
\eeqa
where we have introduced $\delta\equiv\kappa/\ell$. Moreover, 
{}from Eq.(\ref{eq:dV}), we find $r_c=1-\delta$. Putting these into 
Eq.(\ref{eq:V}) gives up to the first order
\beqa
\alpha^2+(\beta+2\kappa)^2+2\delta(\beta+2\kappa)(2\beta+3\kappa+2)
=16(1+2\delta).
\label{eq:ellipse}
\eeqa
Namely, the shadow is now the ellipse suppressed in the $\beta$ direction and elongated in the $\alpha$ direction. If only one of a binary black hole is observed, 
the shape of its shadow determines not only its mass but also the 
information of the other black hole, $\delta=\kappa/\ell$. The other black hole 
should exist in the direction where the shadow is suppressed.

\begin{figure}[htbp]
\includegraphics[width=13cm]{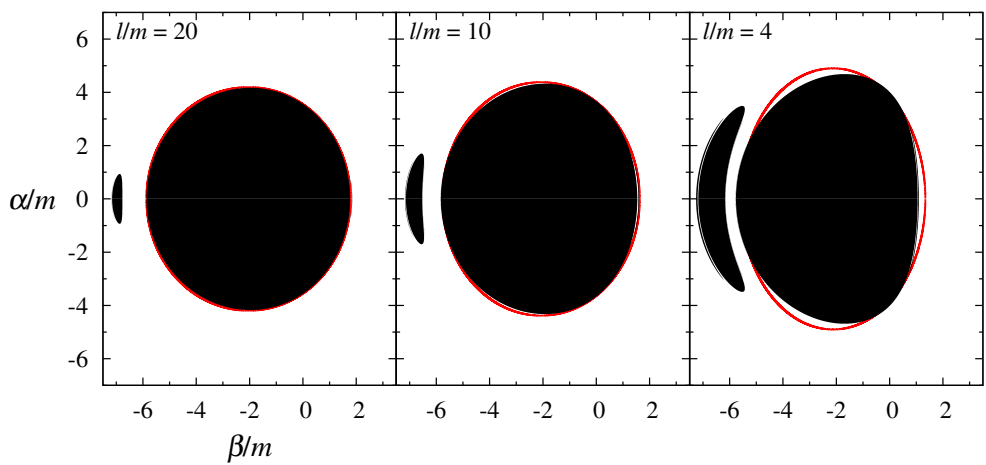}
\caption{The analytic solutions Eq. (\ref{eq:ellipse}) (red) 
superimposed on the black hole shadow for $\ell/m=20,10,4$ with $m=m'$. 
\label{fig:ellipse}}
\end{figure}

In Fig. \ref{fig:ellipse}, we show Eq. (\ref{eq:ellipse}) 
superimposed on the black hole shadows for several $\ell$ with $\kappa=1$. 
We find excellent agreement.

\section{Shadows of Colliding Black Holes: Kastor-Traschen solution}

\begin{figure*}[t]
\begin{center}
\includegraphics[width=11cm,angle=-90]{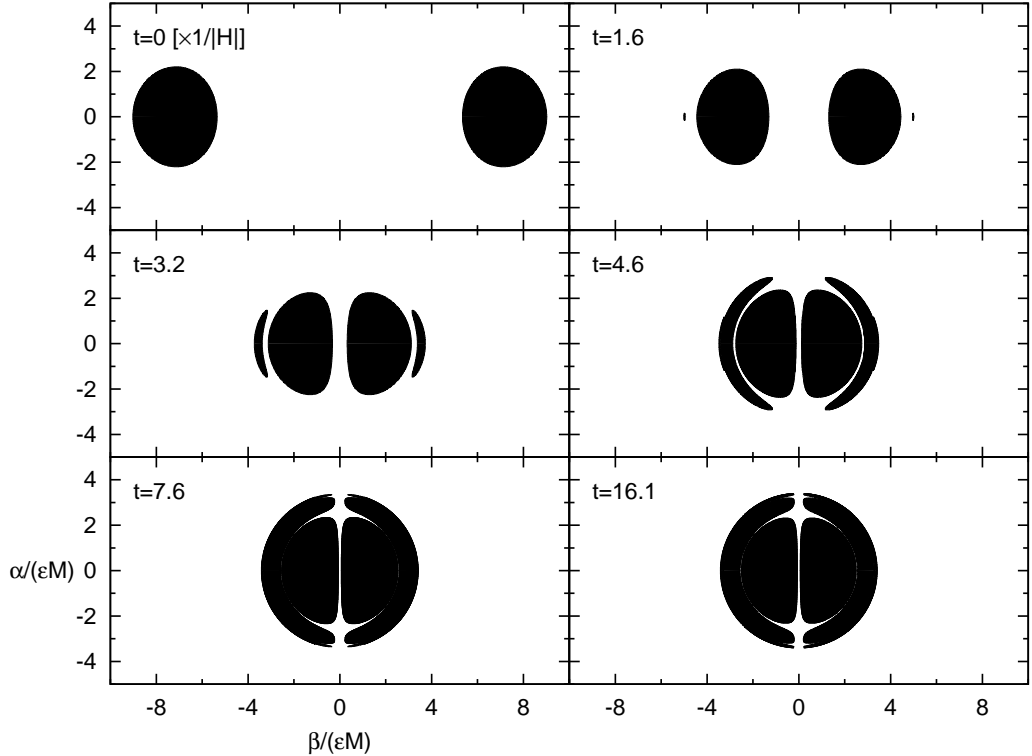}
\end{center}
\caption{The black hole shadows for the two black
hole system in the KT solution plotted in $\alpha$-$\beta$ space 
normalized by $\epsilon M$
with each physical time at the observer 
$t/|H|^{-1}=0,~1.6,~3.2,~4.6,~7.6,~16.1$.  
\label{fig:shadow1}}
\end{figure*}

Finally, we study the shadows of colliding black holes in the Kastor-Traschen(KT) solution \cite{kt}. 
KT solution is a time-dependent generalization of the MP solution and describes 
an arbitrary number of extremely charged black holes in de Sitter universe. 
It is reduced to the Majumdar-Papapetrou(MP) solution when the positive cosmological constant $\Lambda=0$ and is reduced to the extremely charged ($Q=M$) 
Reissner-Nordstr\"om-de Sitter (RNdS) solution.  

The metric in the cosmological coordinate is given by
\begin{eqnarray}
&&ds^2=-a^2\Omega^{-2}d{\tau}^2+a^2\Omega^2(dx^2+dy^2+dz^2),\\
&& a=e^{Ht}=-\frac{1}{H\tau},\quad H=\pm \sqrt{\frac{\Lambda}{3}},\quad\Omega=1+\sum_{i}\frac{m_i}{ar_i},
\label{eq:multi}
\end{eqnarray}
where, $\tau$ and $t$ denote conformal time and physical time respectively.
Here, $H>0$ ($H<0$) corresponds to expansion (contraction). 
In the contracting universe ($H<0$) the KT solution describes the collision of 
black holes.  

Let us consider the situation where an observer is near inside the cosmological
horizon ($r_{obs}\to r_{+}$) in the contracting coordinate. 
The cosmological horizon $r_+$ of $Q=M$ RNdS solution 
in the cosmological coordinate is given by
\begin{eqnarray}
ar_{+}=\frac{1}{2|H|}(1+\sqrt{1-4M|H|})-M. 
\end{eqnarray}
We define the following parameters, which form the celestial coordinate 
system, as
\begin{eqnarray}
\alpha\equiv -\frac{ar_{obs}P^{(\phi)}}{P^{(\tau)}},\quad 
\beta\equiv \frac{ar_{obs}P^{(\theta)}}{P^{(\tau)}},
 \label{eq:ab}
\end{eqnarray}
where $P^{(\mu)}$ are the momenta in the local inertial frame and 
$ar_{obs}$ is the physical distance between the observer and the center of the coordinate.

\subsection{Shadows of Colliding Black Holes}

Let us consider a two black hole system as an example of colliding 
black holes.  Each black hole is located at $z=\pm\ell/2$ in the comoving coordinate.  
We set an observer at a fixed point inside a cosmological horizon in
the physical coordinate. First we take $\theta_{obs}=\pi/2$ 
in terms of the polar coordinate. 

We then numerically calculate the null geodesics from the
observer in the expanding coordinate.  
The null geedesics which eventually fall
into the black hole horizons are regarded as shadows. 
Note that the time reverse is the null geodesics going from the black hole to 
the observer in the contracting universe.

Fig. \ref{fig:shadow1} shows that the shadows of two black holes with
same masses $m_1=m_2$ at each physical time $t$ seen by observers at
$\theta_{obs}=\pi/2$  with $\epsilon\equiv a|H|(r_+-r_{\rm obs})=0.01$.  
We take $M=m_1+m_2=0.1/|H|$.
The separation of two black holes is chosen as {$a \ell=4\times 10^{-3}/|H|$} 
at $t=0$. 
Here, the celestial coordinates $\alpha$
and $\beta$ are normalized by $\epsilon M$ in order to keep the shape of the 
shadows independent of a location of the observer.

At $t=0$ and $t=1.6/| H |$, the black holes are mutually away enough.
However, one can find that their shapes are a little bit elongated in 
the $\alpha$ direction and squeezed in the $\beta$ direction 
from the circles with a radius of 
$4m_i\epsilon/\sqrt{1+4m_i|H|}\sim
1.82\epsilon M$ when they are considered as single black holes in
$\alpha$-$\beta$ space.  {This deformation is caused by the existence of 
the other black hole in the opposite side as explained in Sec.II.D. }

{At $t=3.2/|H|$ (and even at $t=1.6/|H|$)}, 
an eyebrow-like structure around each black hole appears.  
This kind of structure is quite unique to the multi black hole system.
{The reason why these structures appear is explained in Sec.II.C.}    
If the impact parameter of the null geodesics  
is slightly smaller than the radius of the photon sphere, 
these geodesics will eventually
fall into a black hole horizon.  On the other hand, 
for a slightly larger impact parameter, 
the winding geodesics will gradually increase the 
distance to the black hole and eventually go away from the black hole,
or fall into the horizon of the other black hole.  The latter case
creates the eyebrow-like shadow along the main shadow. 
The situation is quite similar to the orbits of the null geodesics 
in the Majumdar-Papapetrou solution.

\begin{figure}[htbp]
\includegraphics[width=8cm,angle=-90]{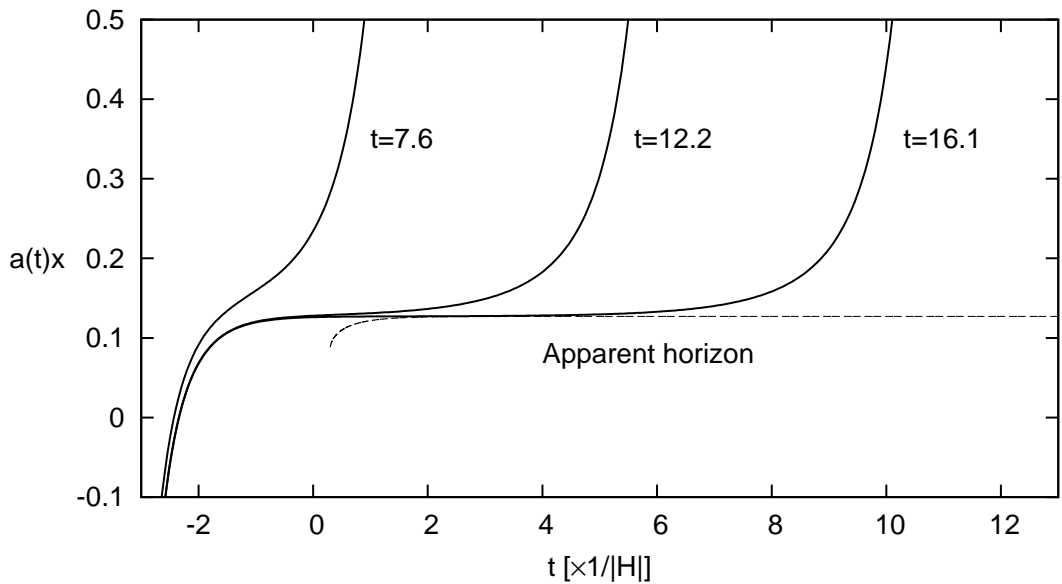}
\caption{The behavior of the null geodesics observed 
in the direction $\alpha=\beta=0$ at $t=7.6,12.2,16.1/|H|$. The dashed curve is 
the evolution of the common apparent horizon enclosing the two black holes. 
The vertical axis is $a(t)x$. The null  geodesics go through the middle of 
 the black holes $(x=0)$ at $t\simeq -2.2/|H|$. 
Later, the black holes get closer and a common apparent horizon appears at $t\simeq 0.3/|H|$. The null geodesics with $t=16.1$ 
stays close to (but just outside) the horizon for a while 
and then reaches the observer when $t=16.1/|H|$. 
\label{fig:t-ar}}
\end{figure}

{At $t=4.6/|H|$ and at $t=7.6/|H|$}, 
the eyebrow-like structures grow and the main shadows
come close each other.  One can find there still remains a region
where photons can go through between the main shadows.   
The reason why such a region remains is the following.  
In a single black hole system, a black hole horizon is enclosed with the 
photon sphere.  
On the other hand, in a two black hole system, two photon spheres
intersect at the $x$-$z$ plane where the null geodesics  
cannot fall into either one of the black holes.  
Accordingly the null geodesics can go through this plane, 
which corresponds to $\beta=0$ in the celestial coordinate 
until two black holes merge and form a horizon. 
%
Even at $t=16.1/|H|$, there still remains a region
where photons can go through between the main shadows. 
\footnote{The merger  of the shadows found in \cite{nitta} is due to the low resolution of the numerical calculation.} 
{According to \cite{nakao}, a common apparent horizon encompassing two black holes appears when $a\ell\siml 10^{-2}/|H|$. 
Since we consider the null geodesics in the contracting universe, 
the geodesics coming into the observer at $t$ 
can go near the black holes at the time much earlier than $t$ when 
the distance between the black holes $a\ell$ is much larger than $0.01/|H|$ so that 
the geodesics can go through in between. 
In fact,  some of the null geodesics observed at $t=16.1/|H|$ pass through the middle 
of the black holes at $t\simeq -2.2/|H|$ when the distance between the black holes 
is $a\ell\simeq 0.036/|H|(> 10^{-2}/|H|)$ and there is no common apparent horizon 
(See Fig. \ref{fig:t-ar}).  }

Overall, the shapes of these shadows look quite similar to those in 
Fig. \ref{fig:MP-shd.pdf}.

\subsection{Shadows of "Coalescing" Black Holes}

\begin{figure}[htbp]
\includegraphics[width=11cm]{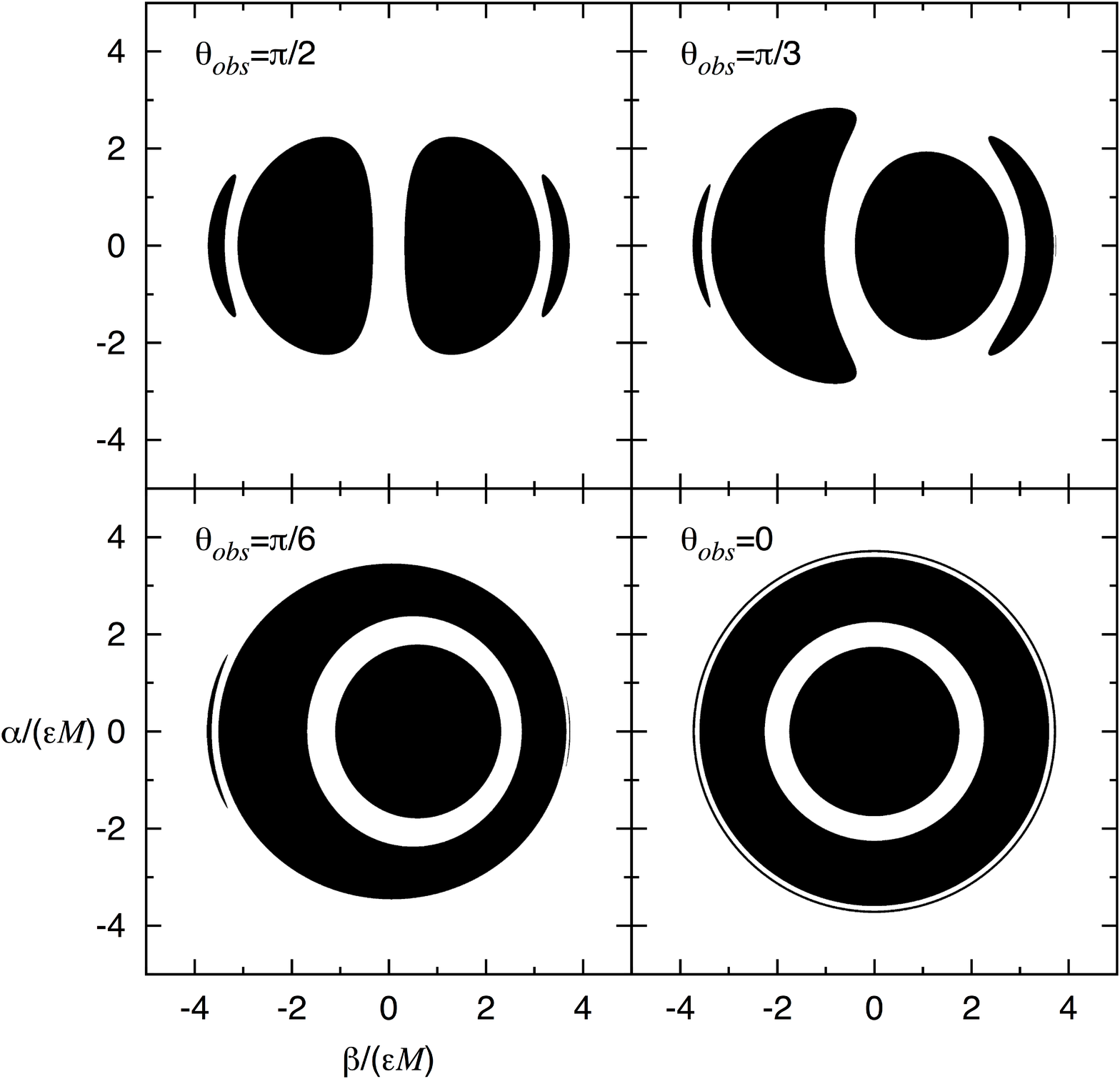}
\caption{The shadows of two black holes in the KT solution at $t=3.7/|H|$ 
viewed by the observer at $\theta_{\rm obs}=\pi/2,\pi/3,\pi/6,0$. 
\label{fig:shadow2}}
\end{figure}

\begin{figure}[htbp]
\includegraphics[width=11cm]{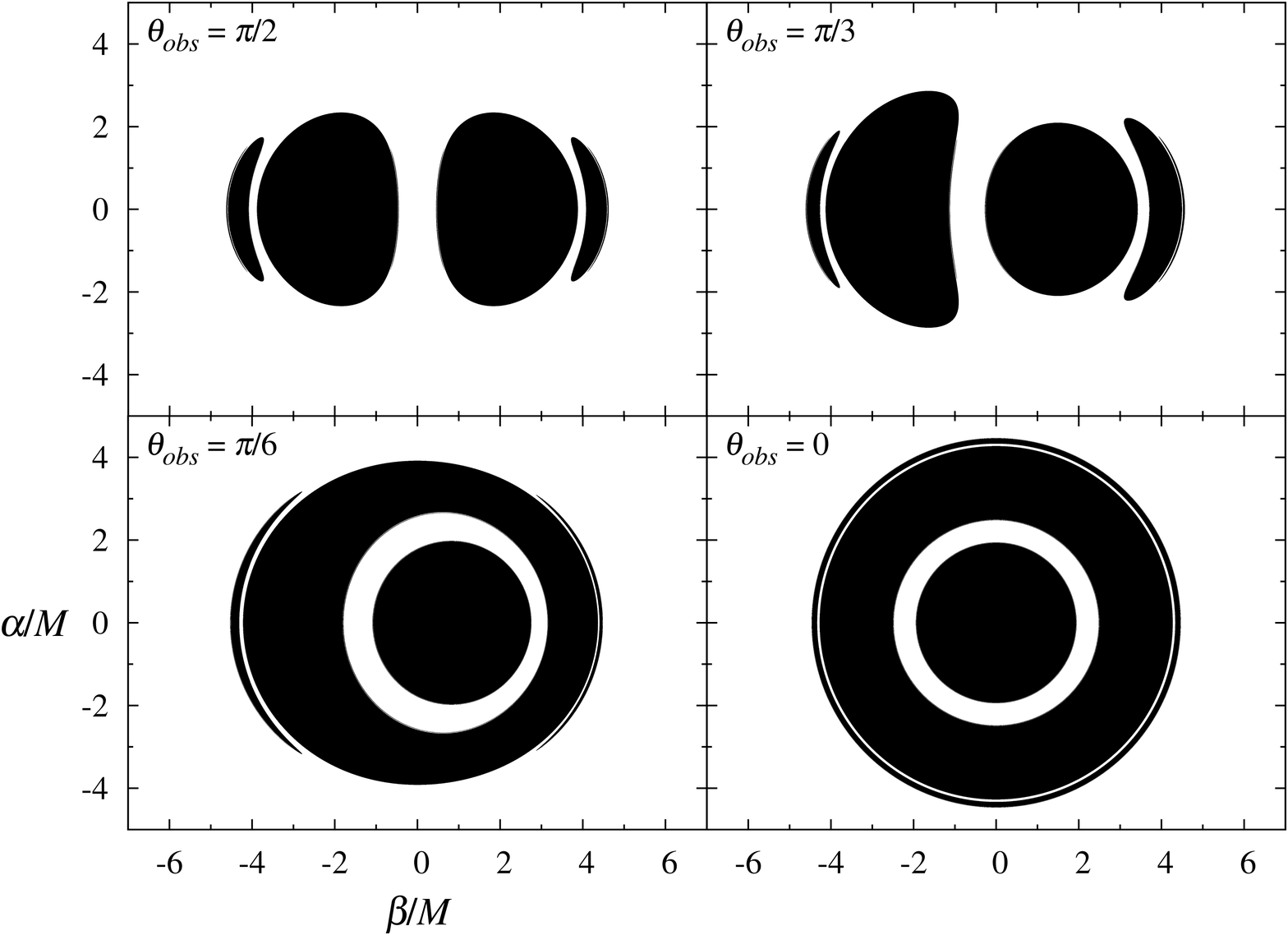}
\caption{The shadows of two black holes in the MP solution with $\ell=2M$ 
viewed by the observer at $\theta_{\rm obs}=\pi/2,\pi/3,\pi/6,0$. 
\label{fig:shadow3}}
\end{figure}

Finally, we consider the situation where one observes black holes from
arbitrary azimuthal directions to mimic the coalescing binaries.  
We have calculated shadows for several
different values of angle $\theta_{obs}$ at $t=3.7$ in Fig. \ref{fig:shadow2}.  
As we decrease
$\theta_{obs}$ from $\pi/2$, the left main shadow of Fig.  \ref{fig:shadow1}  becomes
elongated, and eventually merges with the eyebrow-like structure of the
right side and forms a ring structure surrounding the right main
shadow. For comparison, in Fig. \ref{fig:shadow3}, the shadows of the MP solution 
for  several different angles $\theta_{obs}$ are shown. 
Again, we find that both look similar. 

\section{Summary}

We have studied the null geodesics in the static/dynamic 
multi-black hole solutions: Majumdar-Papapetrou (MP) solution and 
the Kastor-Traschen (KT) solution. We have calculated the shadows of these 
multi-black holes and found that the shadows have structures 
distinct from the mere superposition of the shadow of each black hole: 
 the eyebrow-like structures outside 
the main shadows and the deformation of the shadows. 
We have presented analytic estimates of these structures 
using the MP solution to show that 
the width of the eyebrow is related with the distance between the black holes 
and that the shadows are deformed 
into ellipses due to the presence of the second black holes and the separation between the shadows is larger.

These analytic results help us to have qualitative understanding of the features  
of the shadows of colliding black holes which are studied in our 
previous paper. 
We expect that following two features of black hole shadows are 
general and appear in more realistic situation.  First one is
the eyebrow-like structure which shows up during the merger process.
Second is the deformation of the main shadow and the larger separation than 
the true distance. 
  
These features in the shadows can be used as probes to find
the multi-black hole system at the final stage of its merger process. 
For that purpose, we have presented the shadows of the colliding black holes 
in the KT solution by changing the direction of the observer 
to mimic the coalescence of the binary black holes. 
In order to study the shadows of a realistic black hole binary, the effects of 
the accreation disk should also be considered, which is left for our future study.

\section*{Acknowledgments}

This work was supported in part by a Grant-in-Aid for
Scientific Research from JSPS (No.\,24540287(TC) and
No.\,22340056(NS)) and in part by Nihon University (AY and TC). 
This work was also supported in part by the GCOE Program 
Weaving Science Web beyond Particle-matter Hierarchy at Tohoku University (DN). 
This work
was also supported in part by the Grand-in-Aid for Scientific Research
on Priority Areas No. 467 ``Probing the Dark Energy through an
Extremely Wide and Deep Survey with Subaru Telescope''.
This research has also been supported in part by World Premier
International Research Center Initiative, MEXT, Japan.

\appendix
\section{Derivation of Eq. (\ref{eq:dzeta01})}
\label{app1}
In this Appendix, we show that up to 
${\cal O}(1/\ell)$ $\zeta$ satisfies Eq. (\ref{eq:dzeta01}).

First we prove the following relation
\beqa
|\mu|<\frac{|\zeta|}{r}+{\cal O}(1/\ell).
\label{eq:muzeta}
\eeqa
For convenience, we introduce $\rho\equiv\sqrt{\zeta^2+\xi^2}$. 
First note that since $\zeta^2-\rho^2\mu^2\ge 0$ from Eq.(\ref{eq:dmu}), 
we have $\mu=0$ for $\zeta=0$. 
Therefore, it suffices to prove the relation for $\mu\neq 0$ (or $\zeta\neq 0$). 

{}Since $\mu\to \zeta/r$ for $r\to \infty$, from Eq.(\ref{eq:dmu}) we have for 
$r\to \infty$
\beqa
\frac{d\mu}{dr}=-\frac{\zeta}{r^2\Omega^2}\sqrt{\frac{1-\rho^2\mu^2/\zeta^2}
{1-\rho^2/(r^2\Omega^4)}}.
\label{eq:dmu2}
\eeqa
The sign of the right-hand-side changes at $\rho^2\mu^2=\zeta^2$ or 
$\rho^2=r^2\Omega^4$. The latter corresponds to the turning point of the orbit 
(Eq. (\ref{eq:dr})). Before reaching the turning point, 
there can be several points $r$ 
such that $\rho^2\mu^2=\zeta^2$. Denoting the largest one among such $r$ as $r_1$
we first prove the relation for $r\ge r_1$. 
{ The relation is then consistent for $r <r_1$, because $\mu$ takes a maximum value $\zeta/\rho$ at $r_1$.} 

The function $h$ defined by
\beqa
h\equiv\sqrt{\frac{1-\rho^2\mu^2/\zeta^2}
{1-\rho^2/(r^2\Omega^4)}}
\eeqa
takes the minimum $h=0$ at $r=r_1$ and asymptotes $h=1$ for $r\to\infty$ 
($0\le h < 1$). Using $h$, Eq. (\ref{eq:dmu2}) and Eq. (\ref{eq:dzeta}) 
become 
\beqa
&&\frac{d\mu}{dr}=-\frac{\zeta}{r^2\Omega^2}h,\label{eq:dmu3}\\
&&\frac{d\zeta}{dr}=-2\Omega\psi_{,\mu}h.
\label{eq:dzeta2}
\eeqa
Introducing $g_{\pm}\equiv \mu\pm\frac{\zeta}{r}$, from Eq. (\ref{eq:dmu3}) and 
Eq. (\ref{eq:dzeta2}), $g_{\pm}$ satisfies
\beqa
\frac{dg_{\pm}}{dr}=\frac{\zeta}{r^2}(\pm 1-h/\Omega^2)
\pm 2\frac{\Omega}{r}\psi_{,\mu}h. 
\label{eq:dg}
\eeqa
The second term is 
\beqa
2\frac{\Omega}{r}\psi_{,\mu}h=\frac{2\ell\kappa\Omega}{(r^2+\ell^2-2\ell r\mu)^{3/2}}h>0. 
\eeqa
Then the integration of the second term is 
\beqa
0<A&\equiv &-\int_{\infty}^{r}2\frac{\Omega}{r}\psi_{,\mu}hdr
<-\int_{\infty}^{r}2\frac{\Omega}{r}\psi_{,\mu}dr\nonumber\\
&=&-2\kappa
\int_{\infty}^{r}
\frac{\ell dr}{(r^2+\ell^2)^{3/2}}\left[1+\frac{1}{r}+{\cal O}(1/\ell)\right]
\nonumber\\
&=&\frac{2\kappa}{\ell}\left(1-\frac{r}{\sqrt{r^2+\ell^2}}\right)+{\cal O}(1/\ell^2),\label{eq:A}
\eeqa
which shows that it is the first order quantity. 

Noting $g_{\pm}\to 0$ for $r\to \infty$, the integration of Eq.(\ref{eq:dg}) 
gives
\beqa
(g_{+}+A)(g_{-}-A)=-\left[\int_{\infty}^{r}\frac{\zeta}{r^2}dr\right]^2
+\left[\int_{\infty}^{r}\frac{\zeta}{r^2}\frac{h}{\Omega^2}dr\right]^2
.\label{eq:gp}
\eeqa
Using Eq.(\ref{eq:dzeta2}), the integrals in 
the right-hand-side of Eq.(\ref{eq:gp}) can be rewritten as
\beqa
&&\int_{\infty}^{r}\frac{\zeta}{r^2}dr=
\zeta \int_{\infty}^{r}\frac{1}{r^2}dr
+2\int_{\infty}^{r}\Omega\psi_{,\mu}h
\left[\int_{\infty}^{r}\frac{1}{r^2}dr
\right] dr=-\frac{\zeta}{r}+A,\nonumber\\
&&\int_{\infty}^{r}\frac{\zeta}{r^2}\frac{h}{\Omega^2}dr=
\zeta \int_{\infty}^{r}\frac{h}{r^2\Omega}dr
+2\int_{\infty}^{r}\Omega\psi_{,\mu}h
\left[\int_{\infty}^{r}\frac{h}{r^2\Omega^2}dr
\right] dr.
\label{a10}
\eeqa
The last term in Eq. (\ref{a10}) is less than $A$ because of $0\le h/\Omega^2<1$, therefore this term is the first order quantity.
Substituting these integrals into Eq.(\ref{eq:gp}), we obtain
\beqa
(g_{+}+A)(g_{-}-A)={-\zeta^2\left\{
\left[\int_{\infty}^{r}\frac{1}{r^2}dr\right]^2-
\left[\int_{\infty}^{r}\frac{h}{r^2\Omega^2}dr\right]^2
\right\}+{\cal O}(1/\ell)}. 
\label{eq:gp2}
\eeqa
{}From $0\le h/\Omega^2<1$ we find that the first term in right-hand-side 
 of Eq. (\ref{eq:gp2}) is negative. Therefore, since $A\sim {\cal O}(1/\ell)$, 
\beqa
g_{+}g_{-}<{\cal O}(1/\ell).
\eeqa
Hence we have
\beqa
|\mu|<\frac{|\zeta|}{r}+{\cal O}(1/\ell),
\eeqa
{the relation holds at least up to the turning point of the orbit. }

Next, we show using the relation (\ref{eq:muzeta}) that 
Eq. (\ref{eq:dzeta2}) can be simplified to give Eq. (\ref{eq:dzeta01}). 
Using the relation (\ref{eq:muzeta}) and Eq. (\ref{eq:dzeta2}), we have the inequality for {$r>\rho$}
\beqa
-2\Omega\psi_{,\mu}\sqrt{\frac{1}{1-\rho/(r^2\Omega^2)}}<
\frac{d\zeta}{dr}<-2\Omega\psi_{,\mu}\sqrt{\frac{1-\rho^2/r^2}{1-\rho^2/(r^2\Omega^2)}}.
\eeqa
{We denote $r$ which satisfies $r=\rho(r)$ as $r_{\rho}$. Note that 
$\rho\equiv\sqrt{\zeta^2+\xi^2}\sim {\cal O}(1)$}.
Since the denominator in the square root at $r=\rho$ is $1-1/\Omega^2\sim {\cal O}(1)$,
{there exists constants $c_1$ and $c_2$ of ${\cal O}(1)$,  for $0<r_{\rho}/r\le 1$}
so that 
\beqa
 1+ c_2(r_{\rho}/r)^2\le\sqrt{\frac{1}{1-\rho^2/(r^2\Omega^2)}},\quad
\sqrt{\frac{1-\rho^2/r^2}{1-\rho^2/(r^2\Omega^2)}}\le 1+ c_1(r_{\rho}/r)^2. 
\eeqa
Then we show that the second term in the integral
\beqa
-2\int_{\infty}^{r}\Omega\psi_{,\mu}(1+c_i(r_{\rho}/r)^2),\quad i=1,2,
\eeqa
is of the second order. 

The second term becomes, neglecting higher order terms, 
\beqa
-2\int_{\infty}^{r}\Omega\psi_{,\mu}(r_{\rho}/r)^2dr&=&
-2\kappa \int_{\infty}^{r}\frac{\ell r}{(r^2+\ell^2)^{3/2}}(1+1/r)(r_{\rho}/r)^2dr
\nonumber\\
&=&-2\kappa \frac{r_{\rho}^2}{\ell^2}\int_{\pi/2}^{\chi}\sin\chi(1+\cot\chi/\ell)\cot^2\chi d\chi
\simeq -2\kappa \frac{r_{\rho}^2}{\ell^2}\ln \frac{r}{2\ell},
\eeqa
where we have introduced $r=\ell\tan\chi$. Hence, the integral is of order 
${\cal O}(1/\ell^2)$. 

For $r<r_{\rho}$, from Eq. (\ref{eq:dzeta2}), we have
\beqa
\zeta(r)-\zeta(r_{\rho})\sim {\cal O}(1/\ell^2). 
\eeqa
Therefore, up to ${\cal O}(1/\ell)$,  Eq. (\ref{eq:dzeta2}) is simplified 
by setting $h=1$
\beqa
\frac{d\zeta}{dr}=-2\Omega\psi_{,\mu}.
\nonumber
\eeqa
This is Eq. (\ref{eq:dzeta01}).



\end{document}